\documentclass[aps,prl,twocolumn,showpacs]{revtex4}
\usepackage{amsmath}

\usepackage{graphicx}

\begin{document}

\title{Berry phase and de Haas - van Alphen effect in
 LaRhIn$_5$}

\author{G.~P.~Mikitik}

\author{Yu.~V.~Sharlai}

\affiliation{B.~Verkin Institute for Low Temperature Physics \&
  Engineering, Ukrainian Academy of Sciences,
   Kharkov 61103, Ukraine}

\date{\today}

\begin{abstract}
We explain the experimental data on the magnetization of ${\rm
LaRhIn}_5$ recently published by R. G. Goodrich et al.\  in Phys.\
Rev.\ Lett.\ {\bf 89}, 026401 (2002). We show that the
magnetization of a small electron group associated with a
band-contact line was detected in that paper. These data provide
the first observation of the Berry phase of electrons in metals
via the de Haas - van Alphen effect.
\end{abstract}

\pacs{71.18.+y, 03.65.Vf}

\maketitle

In recent years the concept of the so-called Berry phase \cite{1}
has attracted considerable attention thanks to its fundamental
origin, see, e.g., Refs.~\onlinecite{3,4} and citation therein.
According to Berry, if a Hamiltonian of a quantum system depends
on parameters, and if the parameters undergo adiabatic changes so
that they eventually return to their original values, the wave
function of the system can acquire the so-called geometrical phase
in addition to the familiar dynamical one. This additional phase
(the Berry phase) differs from zero when the trajectory ${\Gamma
}$ of the system in the parameter space is located near a {\it
point} at which the states of the system are degenerate \cite{1}.
In analyzing this situation, Berry assumed that the Hamiltonian of
the system is a Hermitian matrix which is linear in deviations of
the parameters from the point, and he presented his final result
in the pictorial form. He found that such the point can be
considered as a "monopole" in the parameter space when the
geometrical phase is calculated. In other words, the point
"generates" a field which coincides in the form with that of the
monopole, and the flux of this Berry field through the contour
${\Gamma }$ gives the geometrical phase of the system. Evidence
for this phase was obtained in experiments with various physical
systems \cite {3,4}. However, an experimental observation of the
Berry phase for electrons in crystals has proved a challenging
problem (some progress in this direction was achieved only
recently \cite{6,7a,7}).

It is well known (see, e.g., Ref.~\onlinecite{12} ) that the
semiclassical motion of an electron in a crystal in the magnetic
field $H$ can be represented as the motion of the wave vector
${\bf k}$ in an orbit in the Brillouin zone. This orbit is the
intersection of the constant-energy surface, $\varepsilon({\bf
k})=$const, with the plane, $k_z=$const, where ${z}$ is the
direction of the magnetic field ${\bf H}$ and $\varepsilon({\bf
k})$ is electron dispersion relation in the crystal. Berry's
result is applicable to such the electron, with the Brillouin zone
playing the role of the parameter space \cite{Zak}. However, in
crystals with the inversion symmetry and a weak spin-orbit
interaction, the Berry phase of the electrons has the specific
features \cite{prl} which are due to the fact that the electron
states are invariant under the simultaneous inversion of time and
spatial coordinates. This invariance permits one to transform the
Hermitian Hamiltonian of the electron into the real form for any
point of the Brillouin zone. As a consequence, the character of
the energy-band degeneracy differs from that considered by Berry.
Now the electron energy bands $\varepsilon_l({\bf k})$ contact
along {\it lines} in the Brillouin zone, and the lines need not be
symmetry axes \cite{H}. In other words, the degeneracy is not
lifted along these lines, and the monopole in the ${\bf k}$ space
disappears. As it was shown in our paper \cite{prl}, in such the
situation the band-contact lines play the role of infinitely thin
"solenoids" which generate the Berry field with the flux $\pm
\pi$. Although this field is zero outside the solenoids, but if
the electron orbit surrounds the line, the flux threads the orbit,
and the electron acquires the Berry phase $\phi_B=\pm \pi$ when it
moves around this line. It is clear that in this case the Berry
phase does not depend on the shape and the size of the electron
orbit but is specified only by its topological characteristics
(there is either a linking of the orbit with the band-contact line or
there is not).

The Berry phase of the electron modifies \cite{prl} the well-known
semiclassical quantization rule \cite{Sh} for the electron energy
in the magnetic field, $\varepsilon$,
\begin{equation}\label{1}
S(\varepsilon ,k_z)=\frac{2\pi e H}{\hbar c}(n+\gamma ),
\end{equation}
where ${S}$ is the cross-sectional area of the closed electron
orbit in the ${\bf k}$ space; ${n}$ is a large integer (${n>0}$);
${e}$ is the absolute value of the electron charge, and the
constant $\gamma$ is given by
\begin{equation}\label{1a}
\gamma = \frac{1}{2} -\frac{\phi_B}{2\pi}\ .
\end{equation}
The meaning of formula (\ref{1a}) is the following: When
the electron
makes a complete circuit in its orbit, the change of the phase of
its wave function consists of the usual semiclassical part $\hbar
cS/eH$, the shift $-\pi$ associated with the so-called turning
points of the orbit where the semiclassical approximation fails,
and the Berry phase. Equating this change to $2\pi n$, one arrives
at Eqs.~(\ref{1}), (\ref{1a}). Thus, when the electron orbit links
to the band-contact line, one obtains $\gamma=0$ (the values
$\gamma=0$ and $\gamma=1$ are equivalent) instead of the usual
value \cite{Sh} $\gamma=1/2$. This change of $\gamma$ has to
manifest itself in de Haas - van Alphen effect \cite{prl}.

In the recent experimental investigation \cite{UM} of the de Haas
- van Alphen effect in ${\rm LaRhIn}_5$, the oscillations of
magnetization associated with a small cross-section of the Fermi
surface of this metal were detected. The electron cyclotron mass
$m^*$ corresponding to this cross section was also small as
compared with electron mass $m$, $|m^*|\approx 0.067m$. Authors of
that paper attributed these oscillations to a small electron
pocket of the Fermi surface. Besides, in that paper the
magnetization of the electrons of the pocket was studied in the
ultraquantum limit, and the following intriguing contradiction
between the obtained experimental data and the existing theory was
discovered: Since in the ultraquantum limit the electrons occupy
only the lowest Landau level, they have to migrate into large
sheets of the Fermi surface when this level is raised above the
Fermi energy. Hence, the magnetization $M$ of the electrons of the
pocket has to vanish with increasing magnetic field $H$. However,
the experimental data \cite{UM} reveal a finite contribution to
the magnetization even in such magnetic fields.

In this paper we  resolve this contradiction. We show that in fact
a small electron group associated with a band-contact line was
detected in Ref.~\onlinecite{UM}, and the results of
Ref.~\onlinecite{UM} provide the first observation of the Berry
phase of the electrons via the de Haas - van Alphen effect.

A small cross-section of a Fermi surface appears near that point
${\bf k}_0$ of the Brillouin zone for which the two conditions are
fulfilled: Topology of this surface changes at the ${\bf k}_0$ if
the Fermi energy $\varepsilon_F$ is shifted past some critical
energy $\varepsilon_0$, and this $\varepsilon_0$ is close to the
initial Fermi level of the crystal. In the case of the degeneracy
of two electron energy bands of the crystal [say
$\varepsilon_+({\bf k})$ and $\varepsilon_-({\bf k})$] along a
line in the Brillouin zone, these bands near the ${\bf k}_0$
always can be represented in the form \cite{M}:
\begin{equation}\label{2}
\varepsilon_{\pm}({\bf k})\!=\!\varepsilon_0\!+\frac {\hbar^2
k_3^2}{2m_3}+\hbar ({\bf v}_{\perp}\!\!\cdot {\bf k}) \pm \sqrt{
\left (\hbar V_1 k_1 \right )^2\!+\! \left ( \hbar V_2 k_2 \right
)^2},
\end{equation}
where the wave vector ${\bf k}$ is measured from the ${\bf k}_0$;
the constants $m_3$, $V_1$, $V_2$, and ${\bf v}_{\perp}=
(v_1,v_2,0)$ are some parameters of the spectrum; the $k_3$ axis
coincides with the tangent to the band-contact line at the point
${\bf k}_0$, and the ${\bf k}_0$ is defined by the condition that
the band energies in the line [ $\varepsilon_+({\bf
k})=\varepsilon_-({\bf k})$ there ] reach the extremal value
$\varepsilon_0$ at this point. A small extremal cross section can
exist only under the condition, $a_{\perp}^2 \equiv
(v_1/V_1)^2+(v_2/V_2)^2 <1$, which we imply to hold below.

If ${\rm sign}(m_3) (\varepsilon_F- \varepsilon_0)<0$, the Fermi
surface has the shape of a neck, with the band-contact line being
inside the neck, Fig.~1b. Here ${\rm sign}(z)=1$ for $z>0$, and
${\rm sign}(z)=-1$ if $z<0$. As the Fermi energy passes the
critical energy, ${\rm sign}(m_3) (\varepsilon_F-
\varepsilon_0)>0$, the neck is broken, and immediately a new
pocket appears, i.e., the Fermi surface takes the self
intersecting shape, with the band-contact line still lying inside
it, Fig.~1a. Thus, in the case of the degeneracy of the bands, the
${\bf k}_0$ is the point where a self intersecting Fermi surface
appears (or disappears) at $\varepsilon_F = \varepsilon_0$.

\begin{figure}  
\includegraphics[scale=0.95]{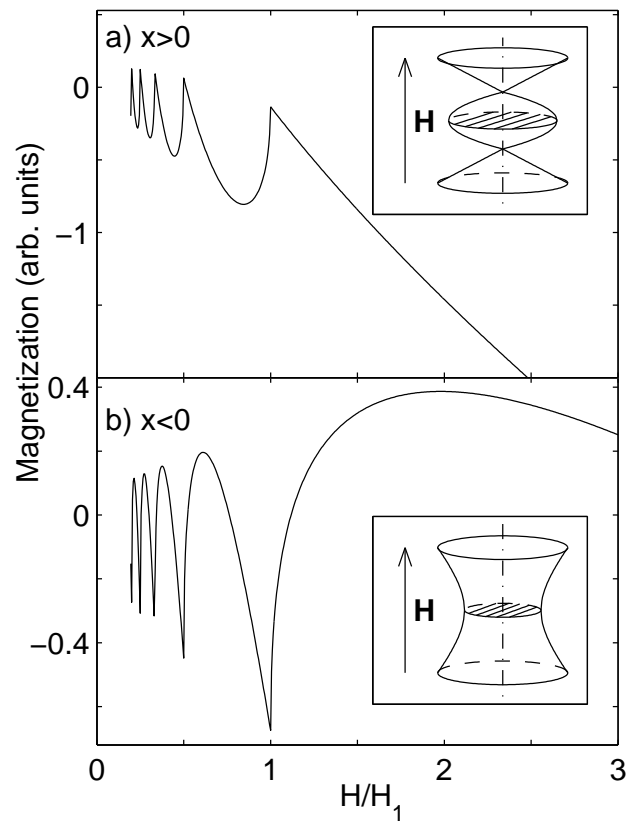}
\caption{\label{fig3} The electron magnetization associated with
the band contact-line, Eq.~(\ref{3.6}), for $x>0$ (a) and for
$x<0$ (b). A sign of $x$ coincides with the sign of
$(\varepsilon_F-\varepsilon_0)m_3$. The electron spectrum is
described by Eq.~(\ref{2}) with $a_{\perp}^2 \equiv
(v_1/V_1)^2+(v_2/V_2)^2 <1$. The field $H_1$ is given by
Eq.~(\ref{3.5}). The insets show the appropriate Fermi surfaces
and the extremal cross sections; the dash-dot lines depict the
band-contact line.
 } \end{figure}  

Let the magnetic field $H$ be in the $k_3$ direction. In this case
the exact Landau levels were found in a vicinity of the critical
energy $\varepsilon_0$ \cite{fnt}. Interestingly, these exact
levels obtained from the appropriate Schrodinger equation coincide
with the semiclassical levels, given by Eq.~(\ref{1}) and
$\gamma=0$, at {\it all} $n$ (even at $n\sim 1$), and not just for
$n\gg 1$.
On the basis of this spectrum, the magnetization $M$
of the electrons with spectrum (\ref{2}) was calculated for an
{\it arbitrary} strength of $H$ \cite{fnt}:
\begin{equation}\label{3.6}
M\!\!= -\!\left (\frac ec \right )^{2}\! \frac { 2^{3/2}
|m_3(\varepsilon_F - \varepsilon_0)|^{1/2}} {\pi^2 \hbar |m^*|}
H_1^{1/4} H^{3/4} f(x),
\end{equation}
where
\[
x={\rm sign}[m_3(\varepsilon_F-\varepsilon_0)]\left ( H_1/H \right
)^{1/2},
\]
and the universal function $f(x)$ is completely independent of the
spectrum,
\begin{equation}\label{3.7}
f(x)=\frac 14 \int_{-x}^\infty dt \left ( \frac 12 - \{ t^2 \}
\right ) {\rm sign} (t) \frac {7t+6x} {\sqrt{x+t}}.
\end{equation}
Here $\{ z\}$ means the fractional part of the number $z$. The
$H_1$ is one of the fields given by the formula,
\begin{equation}\label{3.5}
\frac{e}{ c\hbar} H_n=\frac{S_{ex}}{2\pi}\frac 1 n= \frac
{m^{*}}{2\hbar^2} \frac {(\varepsilon_F - \varepsilon_0)}n ,
\end{equation}
where $n=1,2,...\,$; $S_{ex}$ is the area of the extremal cross
section of the Fermi surface, see Fig.~1, and the cyclotron mass
$m^*$ is proportional to $\varepsilon_F-\varepsilon_0$,
\begin{equation}\label{3.4}
m^{*}= \frac{\varepsilon_F - \varepsilon_0}{
 V_1V_2(1-a_{\perp}^2)^{3/2}},
 \end{equation}
and is small as compared to the electron mass $m$ at
$|\varepsilon_F - \varepsilon_0|\ll m V_1V_2\sim 1-10$eV. The
meaning of the fields $H_n$ will become clear from the subsequent
analysis. Note that we consider $M(H)$ at a fixed $\varepsilon_F$
since in a metal, large sheets of its Fermi surface provide so
large density of states that $\varepsilon_F$ is practically
independent of the magnetic field.

If $|x| \gg 1$, the magnetization (\ref{3.6}) splits into the
oscillation part, that completely agrees with the well-known
Lifshits-Kosevich formula \cite{LK,Sh}, and the smooth
contribution $\chi H$, with the magnetic susceptibility $\chi$
coinciding with that of Ref.\onlinecite{M}. At low temperatures,
\begin{equation}\label{2.7}
 \delta \varepsilon_H\equiv \frac{e\hbar H}{|m^{*}|c}
 \gg 2\pi^2 T,
\end{equation}
the oscillations of the magnetization $M$ prevail over the smooth
part, many harmonics in the Lifshits-Kosevich formula are
relevant, and sharp peaks of $M$ occur when the Landau levels
cross the Fermi energy. It is the field $H_n$ defined by
Eq.~(\ref{3.5}) that gives the position of the peak at crossing
$\varepsilon_F$ by the $n$-th Landau level. Note that at $x>0$ the
oscillations of $M$ result from the {\it maximum} cross section of
the pocket with $k_3=0$, and the peaks of the magnetization are
directed {\it upward}, Fig.~1a, while at $x<0$ the oscillations
result from the {\it minimum} cross section of the neck with
$k_3=0$, and the peaks are directed {\it downward}, Fig.~1b.

Formula (\ref{3.5}) provides possibility to find the band-contact
lines in metals, using Shoenberg's procedure \cite{Sh}. Plotting
experimental values of $1/H_n$ versus $n$, one can state that a
band-contact line has been detected if this dependence is
extrapolated to the origin of the coordinate. If the $\gamma$ were
different from zero, the dependence would be extrapolated to
$-\gamma$. We emphasize that since for the electrons near the
point ${\bf k}_0$ the exact spectrum in the magnetic field
coincides with the semiclassical spectrum, formula (\ref{3.5})
defines the positions of the peaks in $M(H)$ not only for large
$n$ but also for $n\sim 1$. Thus, it is sufficient to use several
last oscillations of $M(H)$ in this detection. This enables one to
find $\gamma$ with maximal accuracy. Note also that the
observation of the sharp peaks for the last oscillations ($n\sim
1$) is the most favorable since $\delta \varepsilon_H \sim
|\varepsilon_F - \varepsilon_0|/n$ in Eq.~(\ref{2.7}).

The ultraquantum limit occurs when $|x|\le 1$, i.e., when $H \ge
H_1$. In this limit the magnetization $M(H)$ cannot be decomposed
into the oscillation parts and the term $\chi H$ [e.g.,
$f(x)\approx 0.156$ at $|x|\ll 1$]. In this case the magnetization
is determined by the Landau levels of the lower band
$\varepsilon_-({\bf k})$ which are {\it all} occupied by the
electrons. It is important that at $H\sim H_1$ the magnetization
far exceeds that of the usual small pockets and necks which
contain no band contact line. Indeed, for such the pockets and
necks the appropriate magnitude of $M$ is of the order of the
prefactor before the function $f(x)$ in Eq.~(\ref{3.6}), but $m^*$
in such the situations is not proportional to $\varepsilon_F -
\varepsilon_0$ and generally is not small, $m^*\sim m$. Moreover,
at $H\sim H_1$ the $M$ in Eq.~(\ref{3.6}) generally {\it is not
small} as compared with the smooth part of the magnetization,
$\chi_0 H_1$, caused by the {\it large electron groups} in metals.
Estimating the smooth part of the magnetic susceptibility of the
large electron groups, $\chi_0$, by the Landau formula for the
electron gas \cite{Sh}, $\chi_0 \sim
(e/c)^2(\varepsilon_F)^{1/2}/\hbar \sqrt m$, we find that at
$H\sim H_1$,
 \[
 \frac{M(H)}{\chi_0 H} \sim \frac{mV_1V_2}{(\varepsilon_F|
 \varepsilon_F -  \varepsilon_0|)^{1/2}}\gg 1.
 \]
In the last inequality we have assumed that $mV_1V_2$ and
$\varepsilon_F$ are of the order of the characteristic energies in
metals, $1-10$eV.

 \begin{figure}  
\includegraphics[scale=0.95]{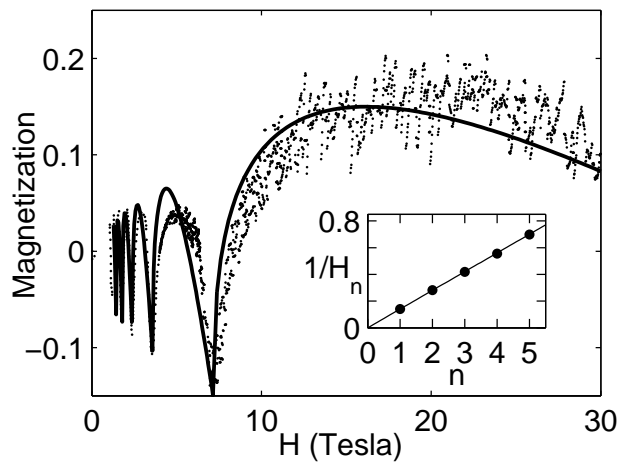}
\caption{\label{fig4} The experimental data \cite{UM} on the
magnetization of ${\rm LaRhIn}_5$ (dots) and the magnetization
calculated from Eqs.~(\ref{3.6}), (\ref{4.1}) (solid line) for
${\rm sign}(m_3)(\varepsilon_F -\varepsilon_0)=-25 meV$,
$|m^*|=0.067m$ ($H_1=7 T$), and $\chi_0$ given in the text. The
inset shows the dependence of the experimental values of $1/H_n$
\cite{UM} on $n$. This dependence gives $\gamma=0$.
 } \end{figure}  

Above we have neglected the spin-orbit interaction in the crystal.
With this interaction, the semiclassical quantization rule
(\ref{1}) is modifies as follows \cite{Sh}:
\begin{equation}\label{s1}
S(\varepsilon ,k_z)=\frac{2\pi e H}{\hbar c}\left ( n+\gamma \pm
\frac {{\rm g} m^{*}} {4m} \right ),
\end{equation}
where $\gamma=1/2$, and $g$ is the so-called $g$ factor of the
electron orbit. Besides, the spin-orbit interaction generally
lifts the degeneracy of the bands $\varepsilon_+({\bf k})$ and
$\varepsilon_-({\bf k})$. But if this interaction is not too
strong so that the gap between these bands is essentially smaller
than the energy gaps between $\varepsilon_{\pm}({\bf k})$ and
other bands of the crystal, the concept of the band-contact line
is still valid approximately. As it was shown in our paper
\cite{zhetf}, if the semiclassical electron orbit in the magnetic
field surrounds such the band-contact line, one has $g\approx
2m/m^*$, and formula (\ref{s1}) is equivalent to Eq.~(\ref{1})
with $\gamma=0$ for all $n$. In other words, equation (\ref{3.5})
for the peak positions is robust to ``switching on'' the
spin-orbit interaction. Note that the $g$ factor is {\it large
even for a very weak spin-orbit interaction}, and this result is
the other manifestation of the nonzero Berry phase (instead of
$\gamma=0$). The spin-orbit interaction also modifies formula
(\ref{3.6}) \cite{fnt}. However, if the splitting of the bands
$\varepsilon_{\pm}({\bf k})$ is small near the point ${\bf k}_0$,
the modification is negligible, and it increases rather {\it
slowly} with the strength of the spin-orbit interaction.

We now apply the above results to the experimental data of
Ref.~\onlinecite{UM}. These data obtained at a low temperature
($1.5$ ${\rm K}$) reveal the sharp peaks in the magnetization of
${\rm LaRhIn}_5$ when the magnetic field $H$ is parallel to the
$[001]$ direction of this tetragonal compound, Fig.~2. The
analysis of the peak positions, see the inset in Fig.~2, gives
$\gamma=0$ \cite{C}, i.e., we conclude that the oscillations in
the magnetization result from some small electron group near the
band-contact line \cite{C1}. Using the experimental value
\cite{UM} of the cyclotron mass, $|m^*|=0.067m$, and the position
of the last peak $H_1\approx 7\, {\rm T}$, we find from
Eq.~(\ref{3.5}) that $|\varepsilon_F -\varepsilon_0|\approx 25
meV$. The downward peaks mean that we deal with the situation
shown in Fig.~1b.

To verify this conclusion, we also compare the theoretical $M(H)$
with the experimental data, Fig.~2. The experimental magnetization
$M_{\rm exp}$ has been approximated by
\begin{equation}\label{4.1}
 M_{\rm exp}(H)=\chi_0 H + M(H) ,
\end{equation}
where $M(H)$ is given by Eq.~(\ref{3.6}), while $\chi_0$ is the
smooth part of the magnetic susceptibility of the large electron
groups in ${\rm LaRhIn}_5$. Thus, we have only the two {\it
constants} to fit the experimental data: the prefactor in
Eq.~(\ref{3.6}), and $\chi_0$. In Fig.~2 we show the theoretical
curve calculated under the condition $\chi_0 H_1/ M(H_1)= 0.14$.
Note that $M(H_1)$ is noticeably larger than $\chi_0 H_1$. It is
also evident that the curve sufficiently well reproduces the
experimental data even without any corrections to $M$ due to the
spin-orbit interaction.

Although the band structure of ${\rm LaRhIn}_5$ was calculated in
Ref.~\onlinecite{cal}, the data presented in that paper do not
permit one to find the band-contact lines in this crystal. To
locate these lines, it would be well to calculate the bands lying
near $\varepsilon_F$ over the Brillouin zone and to trace the
evolution of these bands with the strength of the spin-orbit
interaction. Such the analysis could also clarify one more point:
It turns out that two small and almost equal cross sections
determine the oscillations of $M$ in ${\rm LaRhIn}_5$ \cite{UM}.
This can occur if the direction of the magnetic field slightly
differ from the $[001]$ axis, and if the directions of the
band-contact lines at the equivalent points ${\bf k}_0$ do not
coincide with this axis. Note that when the magnetic field is
tilted away from the $k_3$ axis, the component $H_3$ has to be
inserted in the above formulas \cite{fnt}. Thus, an experimental
investigation of the angular dependences of the two cross sections
could also assist in clarifying this result of
Ref.~\onlinecite{UM}.

In summary, we resolve the contradiction discovered in
Ref.~\onlinecite{UM}. It turns out that a small neck of the Fermi
surface with {\it the band-contact line inside the neck}, see the
inset in Fig.~1b, was discovered in Ref.~\onlinecite{UM}. In this
case the magnetization in the ultraquantum limit does not vanish,
while the positions of the peaks in the oscillation part of $M(H)$
depend on the nonzero Berry phase for the electron orbits in the
magnetic field. In other words, the results of
Ref.~\onlinecite{UM} provide essentially the first observation of
the Berry phase via the de Haas - van Alphen effect.

{}


\begin{thebibliography}{}
\bibitem{C1} It is possible that this line is almost straight and
parallel to the $[001]$ direction, with these properties of the
line being due to the quasi-two dimensional electron energy
spectrum of the crystal.
\end{thebibliography}
\end{document}